\documentclass[prc,preprint,showpacs,preprintnumbers,amsmath,amssymb]{revtex4}

\usepackage[mathscr]{eucal}
\usepackage{graphicx}
\def\slr#1{\setbox0=\hbox{$#1$}           
   \dimen0=\wd0                                 
   \setbox1=\hbox{/} \dimen1=\wd1               
   \ifdim\dimen0>\dimen1                        
      \rlap{\hbox to \dimen0{\hfil/\hfil}}      
      #1                                        
   \else                                        
      \rlap{\hbox to \dimen1{\hfil$#1$\hfil}}   
      /                                         
   \fi}

\def\ksq{k^2}

\def\mytint#1{\!\int\!\!\frac{d^3\!{#1}}{(2\pi)^3}\,}
\def\gev#1{ GeV${}^{#1}$}
\def\be{\begin{eqnarray}}
\def\ee{\end{eqnarray}}

\renewcommand{\theequation}%
    {\arabic{section}.\arabic{equation}}
\makeatletter \@addtoreset{equation}{section} \makeatother

\begin{document}

\preprint{BCCNT: 02/091/315}

\title{Calculation of Hadronic Excitations of the Quark-Gluon Plasma}

\author{Hu Li}
\author{C.M. Shakin}
 \email[email:]{casbc@cunyvm.cuny.edu}

\affiliation{%
Department of Physics and Center for Nuclear Theory\\
Brooklyn College of the City University of New York\\
Brooklyn, New York 11210
}%

\date{September, 2002}

\begin{abstract}
We present calculations of the spectral functions of various
hadronic current correlators at finite temperature, making use of
the Nambu--Jona-Lasinio (NJL) model and the real-time
finite-temperature formalism. We study the scalar-isoscalar
correlation function in a SU(3)-flavor model, as well as the
pseudoscalar and vector correlation functions. We relate our
analysis to our recent calculations of the properties of mesons
for $T<T_c$, which made use of a generalized NJL model that
includes a covariant model of confinement. Here, we exhibit values
of the spectral functions for a range of values of $T>T_c$, where
it is possible to neglect the effects of confinement. We find
important excitations in the scalar sector corresponding to what
are predominately singlet and octet states. The singlet state,
which is at 550 MeV at $T=1.2\, T_c$, evolves from the $f_0(980)$
which has an energy of about 400 MeV before it disappears from the
spectrum of bound states for $T>0.95\,T_c$. The octet state seen
at $T=1.2\,T_c$, which has a mass of about 1100 MeV, evolves from
a nodeless state that has an energy of about 1470 MeV at $T=0$ in
our model. As noted in the literature, these modes may play an
important role in the cooling and hadronization of quark-gluon
droplets excited in heavy-ion collisions. In the past, those
researchers who have used the SU(2)-flavor version of the NJL
(without confinement) and the imaginary-time formalism have found
bound states for the sigma and pion for temperatures that are
larger than $T_c$. In those works it is seen that the sigma and
pion become degenerate after the restoration of chiral symmetry,
as that restoration described in the NJL model. The energy of that
mode then increases with increasing temperature. Our model,
however, has several advantages over previous studies. We are able
to describe meson properties below $T_c$, as well as the
confinement-deconfinement transition. Then, using the real-time
formalism, we are able to describe the widths of the excitations
of the quark gluon plasma that evolve from the pion and $f_0(980)$
with increasing temperature.

\end{abstract}


\maketitle

\section{INTRODUCTION}

Some years ago it was suggested that the quark-gluon plasma was a
``confining medium" in the sense that the important excitations
should be color singlets [1]. At about the same time, Hatsuda and
Kunihiro reported a study of ``soft modes" of the quark-gluon
plasma [2]. These authors used a SU(2)-flavor version of the
Nambu--Jona-Lasinio (NJL) model [3] and considered both finite
temperature and density, making use of the Matsubara
imaginary-time formalism [4,5]. (The standard NJL model does not
describe confinement, so that the question arises as to the
modifications of the results obtained in Ref. [2] if a model of
confinement were to be included.) In Ref. [2] it was found that
the sigma meson mass, which lies in the continuum of the model,
begins to drop in energy with increasing temperature, while the
pion energy increases (slowly) with increasing temperature.
Eventually, the sigma and pion become degenerate in energy, with
the resulting mode increasing in energy with increasing
temperature. These results have been confirmed in a very detailed
recent study making use of a field-theoretic approach in the
calculation of the pseudoscalar and scalar correlation functions
at finite temperature [6]. In that work, phenomenological forms
for the nonperturbative features of the gluon propagator were
introduced and the Schwinger-Dyson and Bethe-Salpeter equations
were solved. In many cases, the model of Ref. [6] yields results
that are similar to those obtained when using the NJL model. It is
of interest to compare the results of our present study with those
obtained in Ref. [6] and to note some of the differences between
our SU(3)-flavor analysis and the SU(2)-flavor analysis of Ref.
[6]. We will return to a discussion of the results obtained in our
work and those obtained in Ref. [6] in Section VI.

Recently, we have also seen an extensive effort in the calculation
of hadronic correlation functions using lattice simulation of QCD
at finite temperature [7-9]. These calculations make use of the
maximum entropy method (MEM) which is reviewed in Ref. [10].
Various peaks are seen in the spectral functions for $T<T_c$ which
persist for $T>T_c$. It is only at relatively large values of $T$
that the correlation functions go over to the smooth behavior
expected for a weakly interacting system [7-9].

Although we have introduced a critical temperature for the
purposes of our discussion, we should note that the quark masses
and vacuum condensates do not go to precisely zero at large
temperature, since the current quark mass is always present in the
model. Indeed, the presence of a small current mass,
$m_u^0=0.0055$ GeV, makes for a significant change in the behavior
of $m_u(T)$ with increasing temperature. On the other hand, it is
useful to introduce a value of $T_c$ of the order of 170 MeV to
facilitate the discussion. The significance of that value for the
characterization of the constituent mass of the up (or down) quark
may be inferred from Fig. 1, which exhibits results calculated for
$m_u^0=0.0055$ GeV and $m_s^0=0.120$ GeV.

 \begin{figure}
 \includegraphics[bb=0 0 300 200, angle=0, scale=1.2]{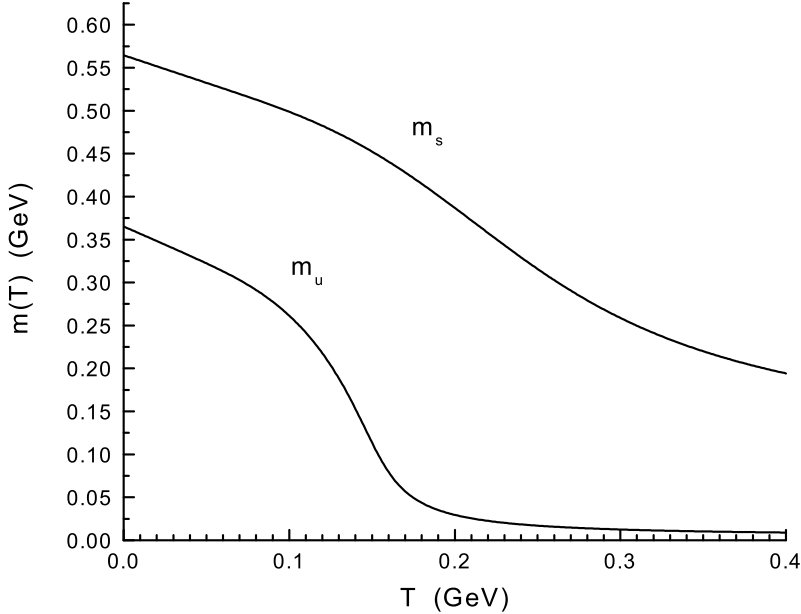}%
 \caption{Temperature-dependent constituent mass values, $m_u(T)$ and $m_s(T)$,
 calculated in a mean-field approximation [12] are shown. [See Eq. (2.2)]. Here $m_u^0=0.0055$ GeV,
 $m_s^0=0.120$ GeV, and $G(T)=5.691[1-0.17(T/T_c)]$\gev2, if we use
 Klevansky's notation [12]. (The value of $G$ used in our work is
 defined such that it is twice the value of $G$ used in Ref. [12].)}
 \end{figure}

It is of interest to obtain insight into the lattice results for
spectral functions by studying hadronic current correlators in a
model, such as the NJL model, that describes the restoration of
chiral symmetry with increasing temperature. We present such a
study in this work. The organization of our work is as follows. In
Section II we review our model of confinement and our calculations
of meson properties for $T<T_c$. In Section III we describe the
calculation of the imaginary part of the vacuum polarization
function of the NJL model, when use is made of the real-time
finite-temperature formalism. Once the imaginary part is
calculated, the real part may be obtained using a dispersion
relation. In Section IV we show how hadronic current correlation
functions may be constructed in terms of the vacuum polarization
functions calculated in Section III. In Section V we present
results of numerical calculations of the various hadronic current
correlation functions considered in this work. Finally, Section VI
contains some further discussion and some comparison of the
Minkowski-space analysis of the present work and the
Euclidean-space analysis of Ref. [6].

\section{calculation of meson properties at finite temperature in a
generalized NJL model with confinement}

It is useful to record the Lagrangian used in our calculations of
meson properties \be {\cal L}=&&\bar q(i\slr
\partial-m^0)q +\frac{G_S}{2}\sum_{i=0}^8[
(\bar q\lambda^iq)^2+(\bar qi\gamma_5 \lambda^iq)^2]\nonumber\\
&&-\frac{G_V}{2}\sum_{i=0}^8[
(\bar q\lambda^i\gamma_\mu q)^2+(\bar q\lambda^i\gamma_5 \gamma_\mu q)^2]\nonumber\\
&& +\frac{G_D}{2}\{\det[\bar q(1+\gamma_5)q]+\det[\bar
q(1-\gamma_5)q]\} \nonumber\\
&&+ {\cal L}_{conf}\,, \ee Here, $m^0$ is a current quark mass
matrix, $m^0=\mbox{diag} \,(m_u^0, \,m_d^0, \,m_s^0)$. The
$\lambda^i$ are the Gell-Mann (flavor) matrices,
$\lambda^0=\sqrt{2/3}\mathbf{\,1}$ with $\mathbf{\,1}$ being the
unit matrix. The fourth term on the right-hand side of Eq. (2.1)
is the 't Hooft interaction. Finally, ${\cal L}_{conf}$ represents
the model of confinement we have used in our work.

As noted earlier, we have recently reported results of our
calculations of the temperature dependence of the spectra of
various mesons [11]. These calculations were made using our
generalized NJL model which includes a covariant model of
confinement. We have presented results for the $\pi$, $K$, $a_0$,
$f_0$ and $K_0^*$ mesons in Ref. [11].

There are three important temperature-dependent features of our
model. Temperature-dependent constituent quark masses were
calculated using the equation [12] \be
m(T)=m^0+2G_S(T)N_c\frac{m(T)}{\pi^2}+\int_0^\Lambda
dp\frac{P^2}{E_p}\tanh(\frac12\beta E_p)\,.\ee Here, $m^0$ is the
current quark mass, $G_S(T)$ is a temperature-dependent coupling
constant introduced in our model [11] \be
G_S(T)=G_S\left[1-0.17\left(\frac T{T_c}\right)\right]\,,\ee
$N_c=3$ is the number of colors, $\beta=1/T$ and $E_p=\left[\vec
p\;{}^2+m^2(T)\right]^{1/2}$. Further, $\Lambda=0.631$ GeV is a
cutoff such that $|\vec p|\leq\Lambda$. Results obtained for the
up (or down) and strange quark masses are given in Fig. 1. In
calculating the constituent mass values we have neglected the
confining interaction. That interaction was taken into account in
our earlier Euclidean-space calculation of the quark self-energy
[13], which also included the effects related to 't Hooft
interaction. We found that, to a good approximation, we could
neglect the confining and 't Hooft interactions, if we modified
the value of the NJL coupling constant, $G_S$, and we adopt that
approach when using Eq. (2.2).

In addition to the temperature dependence of the coupling constant
and constituent mass values, we also introduced a
temperature-dependent confining potential, whose form was
motivated by recent lattice simulations of QCD in which the
temperature dependence of the confining interaction was calculated
with dynamical quarks [14]. (See Fig. 2.) In order to include such
effects, we modified the form of our confining interaction,
$V^C(r)=\kappa r\exp[-\mu r]$, by replacing $\mu$ by \be
\mu(T)=\frac{\mu_0}{\left[1-0.7\left(\displaystyle\frac{T}{T_c}\right)^2\right]}\,,\ee
with $\mu_0=0.010$ GeV. The maximum value of $V^C(r, T)$ is then
\be
V_{max}^C(T)&=&\frac\kappa{\mu(T)e}\,,\\&=&\frac{\kappa\left[1-0.7({T}/{T_c})^2\right]}
{\mu_0e}\,,\ee with $r_{max}=1/\mu(T)$. To better represent the
qualitative features of the results shown in Fig. 2, we can use
$V^C(r, T)=\kappa r\exp[-\mu(T) r]$ for $r\leq r_{max}$ and
$V^C(r, T)=V_{max}^C(T)$ for $r>r_{max}$. We also note that we use
Lorentz-vector confinement and carry out all our calculations in
momentum-space. The value of $\kappa$ used in our work is
0.055\gev2.

\begin{figure}
 \includegraphics[bb=0 0 250 350, angle=-90, scale=1.2]{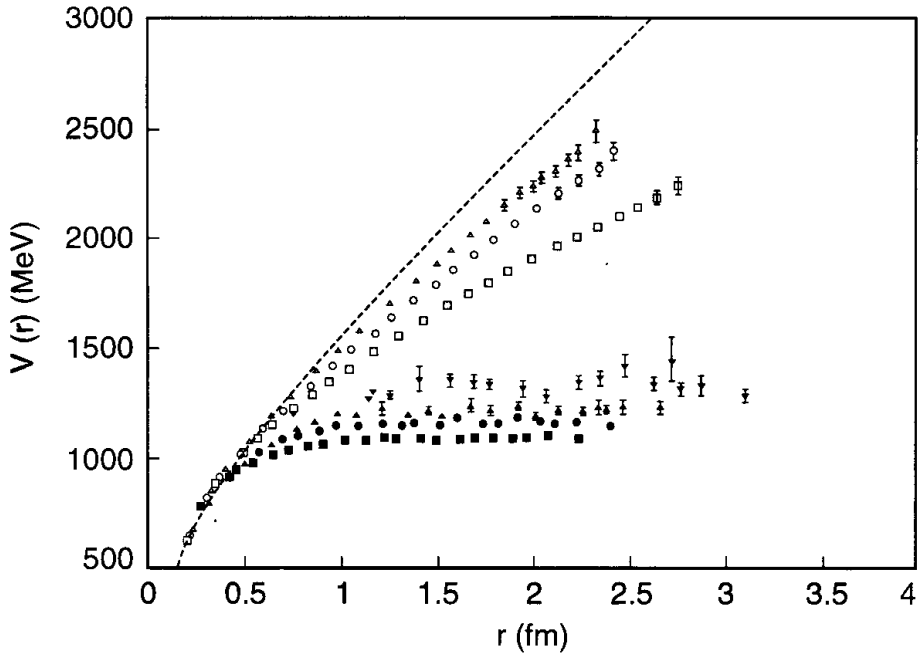}%
 \caption{A comparison of quenched (open symbols) and unquenched (filled symbols) results for
 the interquark potential at finite temperature [14]. The dotted line is the zero temperature
 quenched potential. Here, the symbols for $T=0.80T_c$ [open triangle], $T=0.88T_c$
 [open circle], $T=0.80T_c$ [open square], represent the quenched
 results. The results with dynamical fermions are given at $T=0.68T_c$ [solid downward-pointing
 triangle], $T=0.80T_c$ [solid upward-pointing triangle], $T=0.88T_c$ [solid circle],
 and $T=0.94T_c$ [solid square].}
 \end{figure}

We do not attempt to review the details of our calculations of
temperature-dependent meson spectra [11]. However, in Figs. 3 and
4 we show the results for the $\pi$ and $f_0$ mesons obtained in
Ref. [11]. (These calculations were made with the
temperature-dependent potentials shown in Fig. 5.) In Figs. 3 and
4, it may be seen that, as $T\rightarrow T_c$, fewer states are
bound, with no bound states remaining at $T=T_c$\,. Our confining
potential is finite at $T=T_c$\,, but has a zero value for
$T=1.195T_c$\,. (The potential is defined to be equal to zero for
$T\geq 1.195T_c$\,.) However, it is important to note that
$G_S(T)=0$ only for the quite large value $T=5.88T_c$\,. (The
coupling constants of the model are taken to be zero for
$T\geq5.88T_c$\,.) Even though the confining potential is absent
for $T\geq1.2T_c$\,, the continued presence of the short-range NJL
interaction has important consequences for the dynamics of the
quark-gluon plasma, as we will explore in the following sections.

 \begin{figure}
 \includegraphics[bb=0 0 440 670, angle=0, scale=0.5]{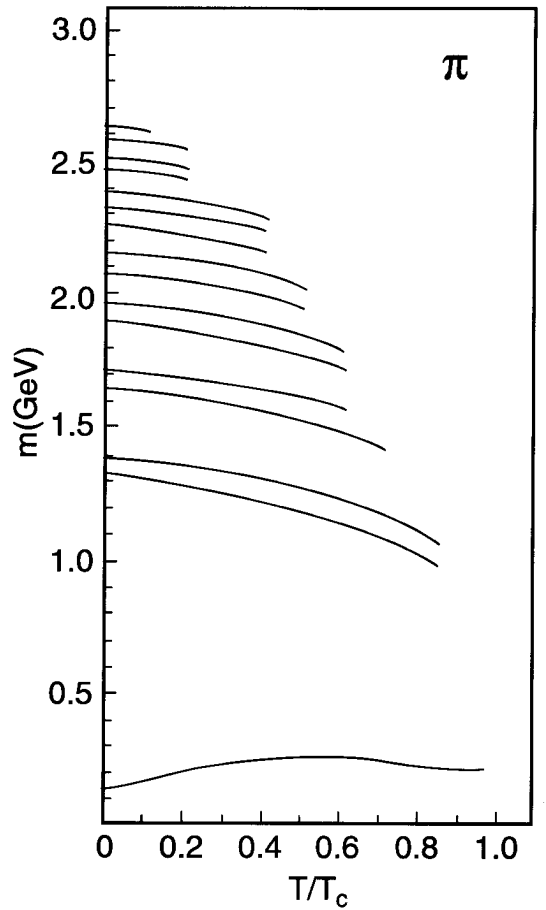}%
 \caption{The mass values of the pionic states calculated in Ref.
 [11] with $G_\pi(T)=13.49[1-0.17\,T/T_c]$\gev{-2}, $G_V(T)=11.46[1-0.17\,T/T_c]$\gev{-2},
 and the quark mass values given in Fig. 1. The value of the pion mass is
 0.223 GeV at $T/T_c=0.90$, where $m_u(T)=0.102$ GeV and $m_s(T)=0.449$ GeV.}
 \end{figure}

 \begin{figure}
 \includegraphics[bb=0 0 440 670, angle=0, scale=0.5]{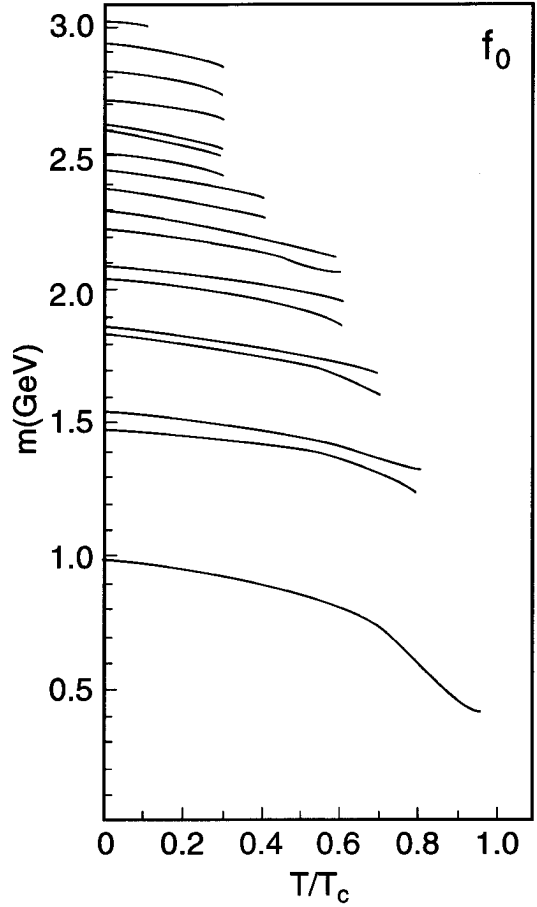}%
 \caption{Mass values of the $f_0$ mesons calculated in Ref. [11] with
 $G_{00}(T)=14.25[1-0.17\,T/T_c]$\gev{-2}, $G_{88}(T)=10.65[1-0.17\,T/T_c]$\gev{-2},
 $G_{08}(T)=0.495[1-0.17\,T/T_c]$\gev{-2}, and $G_{80}(T)=G_{08}(T)$ in a singlet-octet
 representation. The quark mass values used are shown in Fig. 1. The
 $f_0$ has a mass of 0.400 GeV at $T/T_c=0.95$.}
 \end{figure}

 \begin{figure}
 \includegraphics[bb=0 0 300 200, angle=0, scale=1.2]{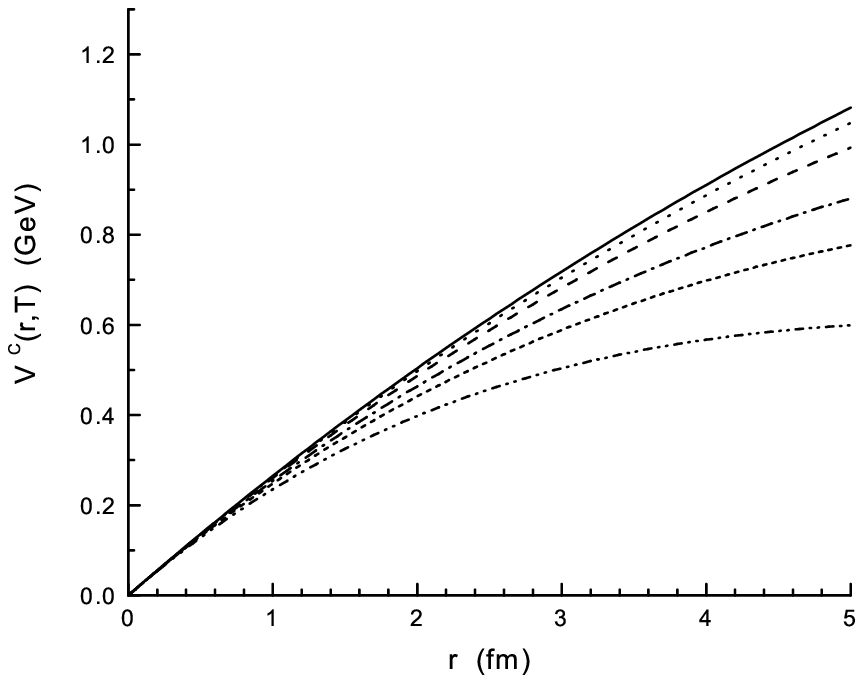}%
 \caption{The potential $V^C(r, T)$ is shown for $T/T_c=0$ [solid line],
 $T/T_c=0.4$ [dotted line], $T/T_c=0.6$ [dashed line], $T/T_c=0.8$ [dashed-dotted line],
 $T/T_c=0.9$ [short dashes], $T/T_c=1.0$ [dashed-(double) dotted line]. Here,
 $V^C(r,T)=\kappa r\exp[-\mu(T)r]$, with $\mu(T)=0.01\mbox{GeV}/[1-0.7(T/T_c)^2]$ and
 $\kappa=0.055$\gev2.}
 \end{figure}

\section{polarization functions at finite temperature}

The basic polarization functions that are calculated in the NJL
model are shown in Fig. 6. We will consider calculations of such
functions in the frame where $\vec P=0$. In our earlier work,
calculations were made after a confinement vertex was included.
That vertex is represented by the filled triangular region in Fig.
6. However, we here consider calculations for $T\geq 1.2T_c$ where
confinement may be neglected. We will, however, use the
temperature-dependent mass values shown in Fig. 1.

 \begin{figure}
 \includegraphics[bb=0 0 600 350, angle=0, scale=0.4]{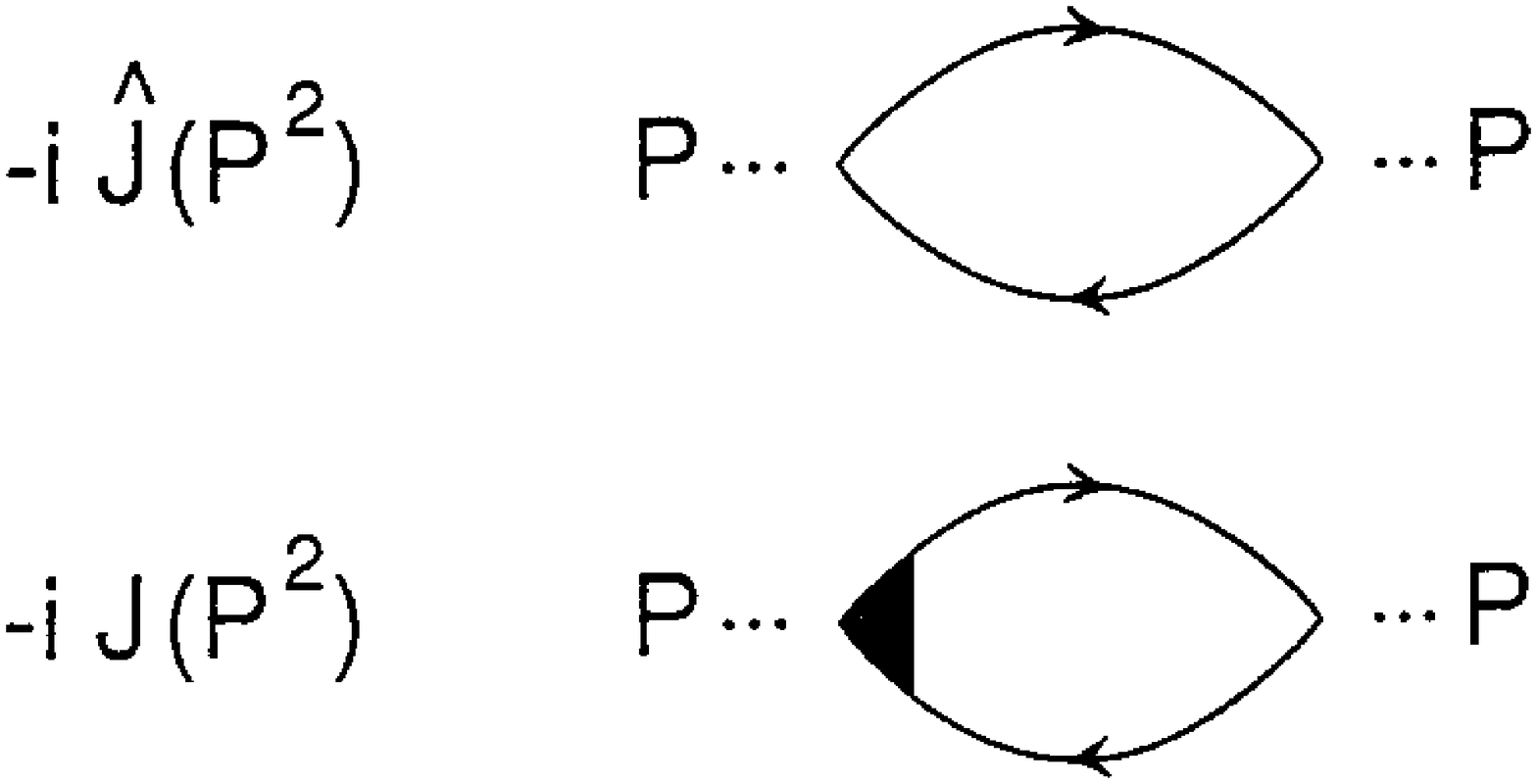}%
 \caption{The upper figure represents the basic polarization diagram of the
 NJL model in which the lines represent a constituent quark and a constituent
 antiquark. The lower figure shows a confinement vertex [filled triangular
 region] used in our earlier work. For the present work we neglect confinement
 for $T\geq1.2T_c$, with $T_c=170$ MeV.}
 \end{figure}

The procedure we adopt is based upon the real-time
finite-temperature formalism, in which the imaginary part of the
polarization function may be calculated. Then, the real part of
the function is obtained using a dispersion relation. The result
we need for this work has been already given in the work of Kobes
and Semenoff [15]. (In Ref. [15] the quark momentum in Fig. 6 is
$k$ and the antiquark momentum is $k-P$. We will adopt that
notation in this section for ease of reference to the results
presented in Ref. [15].) With reference to Eq. (5.4) of Ref. [15],
we write the imaginary part of the scalar polarization function as
\be \mbox{Im}\,J^S(\textit{P}\,{}^2,
T)=\frac12(2N_C)\beta_S\,\epsilon(\textit{P}\,{}^0)\mytint
ke^{-\vec
k\,{}^2/\alpha^2}\left(\frac{2\pi}{2E_1(k)2E_2(k)}\right)\\\nonumber
\{(1-n_1(k)-n_2(k))
\delta(\textit{P}\,{}^0-E_1(k)-E_2(k))\\\nonumber-(n_1(k)-n_2(k))
\delta(\textit{P}\,{}^0+E_1(k)-E_2(k))\\\nonumber-(n_2(k)-n_1(k))
\delta(\textit{P}\,{}^0-E_1(k)+E_2(k))\\\nonumber-(1-n_1(k)-n_2(k))
\delta(\textit{P}\,{}^0+E_1(k)+E_2(k))\}\,.\ee Here, $E_1(k)=[\vec
k\,{}^2+m_1^2(T)]^{1/2}$. Relative to Eq. (5.4) of Ref. [15], we
have changed the sign, removed a factor of $g^2$ and have included
a statistical factor of $2N_C$, where the factor of 2 arises from
the flavor trace. In addition, we have included a Gaussian
regulator, $\exp[-\vec k\,{}^2/\alpha^2]$, with $\alpha=0.605$
GeV, which is the same as that used in most of our applications of
the NJL model in the calculation of meson properties [16-21]. We
also note that \be n_1(k)=\frac1{e^{\,\beta E_1(k)}+1}\,,\ee and
\be n_2(k)=\frac1{e^{\,\beta E_2(k)}+1}\,.\ee For the calculation
of the imaginary part of the polarization function, we may put
$\ksq=m_1^2(T)$ and $(k-P)^2=m_2^2(T)$, since in that calculation
the quark and antiquark are on-mass-shell. In Eq. (3.1) the factor
$\beta_S$ arises from a trace involving Dirac matrices, such that
\be \beta_S&=&-\mbox{Tr}[(\slr k+m_1)(\slr k-\slr P+m_2)]\\
&=&2P^2-2(m_1+m_2)^2\,,\ee where $m_1$ and $m_2$ depend upon
temperature. In the frame where $\vec P=0$, and in the case
$m_1=m_2$, we have $\beta_S=2P_0^2(1-{4m^2}/{P_0^2})$. For the
scalar case, with $m_1=m_2$, we find \be \mbox{Im}\,J^S(P^2,
T)=\frac{N_cP_0^2}{4\pi}\left(1-\frac{4m^2}{P_0^2}\right)^{3/2}
e^{-\vec k\,{}^2/\alpha^2}[1-2n_1(k)]\,,\ee where \be \vec
k\,{}^2=\frac{P_0^2}4-m^2(T)\,.\ee

We may evaluate Eq. (3.6) for $m(T)=m_u(T)=m_d(T)$ and define
$\mbox{Im}\,J_u^S(P^2, T)$. When we put $m(T)=m_s(T)$, we define
$\mbox{Im}\,J_s^S(P^2, T)$. These two functions will be needed for
our calculation of the scalar-isoscalar correlator discussed in
the next section. The real parts of the functions $J_u^S(P^2, T)$
and $J_s^S(P^2, T)$ may be obtained using a dispersion relation,
as noted earlier.

For pseudoscalar mesons, we replace $\beta_S$ by
\be \beta_P&=&-\mbox{Tr}[i\gamma_5(\slr k+m_1)i\gamma_5(\slr k-\slr P+m_2)]\\
&=&2P^2-2(m_1-m_2)^2\,,\ee which for $m_1=m_2$ is $\beta_P=2P_0^2$
in the frame where $\vec P=0$. We find, for the $\pi$ mesons, \be
\mbox{Im}\,J^P(P^2,T)=\frac{N_cP_0^2}{4\pi}\left(1-\frac{4m(T)^2}{P_0^2}\right)^{1/2}
e^{-\vec k\,{}^2/\alpha^2}[1-2n_1(k)]\,,\ee where $ \vec
k\,{}^2={P_0^2}/4-m_u^2(T)$, as above. Thus, we see that, relative
to the scalar case, the phase space factor has an exponent of 1/2
corresponding to a \textit{s}-wave amplitude. For the scalars, the
exponent of the phase-space factor is 3/2, as seen in Eq. (3.6).

For a study of vector mesons we consider \be
\beta_{\mu\nu}^V=\mbox{Tr}[\gamma_\mu(\slr k+m_1)\gamma_\nu(\slr
k-\slr P+m_2)]\,,\ee and calculate \be
g^{\mu\nu}\beta_{\mu\nu}^V=4[P^2-m_1^2-m_2^2+4m_1m_2]\,,\ee which,
in the equal-mass case, is equal to $4P_0^2+8m^2(T)$, when
$m_1=m_2$ and $\vec P=0$. This result will be needed when we
calculate the correlator of vector currents in the next section.
Note that for the elevated temperatures considered in this work
$m_u(T)=m_d(T)$ is quite small, so that $4P_0^2+8m_u^2(T)$ can be
approximated by $4P_0^2$ when we consider the $\rho$ meson.

\section{calculation of hadronic current correlation functions}

In this section we consider the calculation of
temperature-dependent hadronic current correlation functions. The
general form of the correlator is a transform of a time-ordered
product of currents, \be iC(P^2, T)=\int d^4xe^{\,iP\cdot
x}<<\mbox T(j(x)j(0)>>\,,\ee where the double bracket is a
reminder that we are considering the finite temperature case.

For the study of pseudoscalar states, we may consider currents of
the form $j_{P,\,i}(x)=\bar q(x)i\gamma_5\lambda^iq(x)$, where, in
the case of the $\pi$ mesons, $i=1,2,$ and 3. For the study of
scalar-isoscalar mesons, we introduce $j_{S,\,i}(x)=\bar
q(x)\lambda^iq(x)$, where $i=0$ for the flavor-singlet current and
$i=8$ for the flavor-octet current.

In the case of the $\pi$ mesons, the correlator may be expressed
in terms of the basic vacuum polarization function of the NJL
model, $J_P(P^2, T)$ [12, 22, 23]. Thus, \be C_\pi(P^2,T)=J_P(P^2,
T)\frac1{1-G_\pi(T)J_P(P^2, T)}\,,\ee where $G_\pi(T)$ is the
coupling constant appropriate for our study of the $\pi$ mesons.
We have found $G_\pi(0)=13.51$\gev{-2} by fitting the pion mass in
a calculation made at $T=0$, with $m_u=m_d=0.364$ GeV [11]. (We
will describe the calculation of $J_P(P^2, T)$ later in this
section.)

The calculation of the correlator for scalar-isoscalar states is
more complex, since there are both flavor-singlet and flavor-octet
states to consider. We may define polarization functions for $u$,
$d$ and $s$ quarks: $J_u(P^2, T)$, $J_d(P^2, T)$ and $J_s(P^2,
T)$. (These functions do not contain the factor of 2 that would
otherwise appear when forming the flavor trace.) In terms of these
polarization functions we may then define \be J_{00}(P^2,
T)=\frac23[J_u(P^2, T)+J_d(P^2, T)+J_s(P^2, T)]\,,\ee \be
J_{08}(P^2, T)=\frac{\sqrt2}3[J_u(P^2, T)+J_d(P^2, T)-2J_s(P^2,
T)]\,,\ee and \be J_{88}(P^2, T)=\frac13[J_u(P^2, T)+J_d(P^2,
T)+4J_s(P^2, T)]\,.\ee We also introduce the matrices \be J(P^2,
T)=\left[\begin{array}{cc}J_{00}(P^2, T)&J_{08}(P^2,
T)\\J_{80}(P^2, T)&J_{88}(P^2, T)\end{array}\right]\,,\ee \be
G(T)=\left[\begin{array}{cc}G_{00}(T)&G_{08}(T)\\G_{80}(T)&G_{88}
(T)\end{array}\right]\,,\ee and \be C(P^2,
T)=\left[\begin{array}{cc}C_{00}(P^2, T)&C_{08}(P^2,
T)\\C_{80}(P^2, T)&C_{88}(P^2, T)\end{array}\right]\,.\ee We then
write the matrix relation \be C(P^2, T)=J(P^2, T)[1-G(T)J(P^2,
T)]^{-1}\,.\ee

For some purposes it may be useful to also define a \textit{t}
matrix \be t(P^2, T)=[1-G(T)J(P^2, T)]^{-1}G(T)\,,\ee where
$t(P^2, T)$ has the structure shown in Eqs. (4.6)-(4.8). The same
resonant structures are seen in both $C(P^2, T)$ and $t(P^2, T)$.

For a study of the correlators related to the $\rho$ meson, we
introduce conserved vector currents $ j_{\mu,\,i}(x)=\bar
q(x)\gamma_\mu\lambda_iq(x)$ with $i=1,2$ and 3. In this case we
define \be J_\rho^{\mu\nu}(P^2,
T)=\left(g\,{}^{\mu\nu}-\frac{P\,{}^\mu
P\,{}^\nu}{P^2}\right)J_\rho(P^2, T)\ee and \be
C_\rho^{\mu\nu}(P^2, T)=\left(g\,{}^{\mu\nu}-\frac{P\,{}^\mu
P\,{}^\nu}{P^2}\right)C_\rho(P^2, T)\,,\ee taking into account the
fact that the current $j_{\mu,\,i}(x)$ is conserved. We may then
use the fact that \be
J_\rho(P^2,T)&=&\frac13g_{\mu\nu}J_\rho^{\mu\nu}(P^2,T)\\
&=&\frac{2N_c}3\left[\frac{P_0^2+2m_u^2(T)}{4\pi}\right]
\left(1-\frac{4m_u^2(T)}{P_0^2}\right)^{1/2}e^{-\vec
k\,{}^2/\alpha^2}[1-2n_1(k)]\\&\simeq&\frac23J_\pi(P^2,T)\,.\ee
See Eq. (3.7) for the specification of $k=|\vec k|$. We then have
\be C_\rho(P^2,T)=J_\rho(P^2,T)\frac1{1-G_V(T)J_\rho(P^2,T)}\,.\ee

\section{results of numerical calculations}

We begin our presentation with the results we have obtained for
scalar-isoscalar excitations. In Fig. 7 we show the values of
$\mbox{Im}\,J_{00}(P^2)$ for $T/T_c=1.2$, 1.6, 2.0, 4.0 and 6.0.
Figure 8 exhibits similar results for $\mbox{Im}\,J_{08}(P^2)$,
while Fig. 9 exhibits the results obtained for
$\mbox{Im}\,J_{88}(P^2)$. The real parts of these functions are
obtained using a dispersion relation. Once we have the real and
imaginary parts of these functions, we can calculate the elements
of the correlation functions.

 \begin{figure}
 \includegraphics[bb=0 0 280 220, angle=0, scale=1]{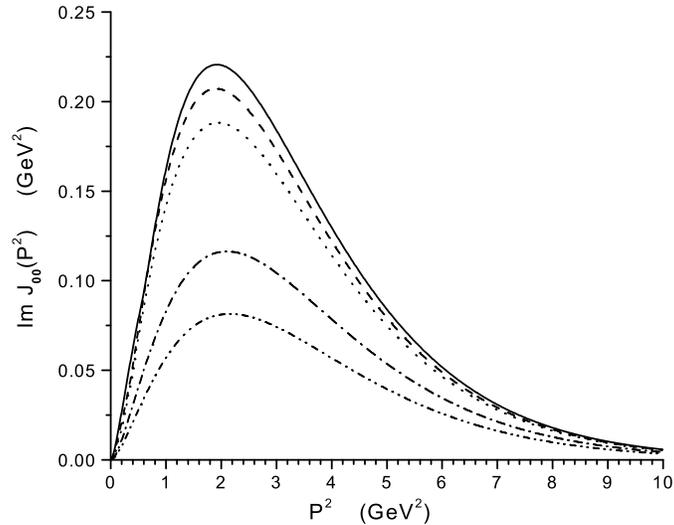}%
 \caption{Values of $\mbox{Im}\,J_{00}(P^2)$ are shown. Here,
 $T/T_c=1.2$ [solid line], 1.6 [dashed line], 2.0 [dotted line],
 4.0 [dashed-dotted line] and 6.0 [dashed-(double) dotted line].}
 \end{figure}

 \begin{figure}
 \includegraphics[bb=0 0 280 230, angle=0, scale=1]{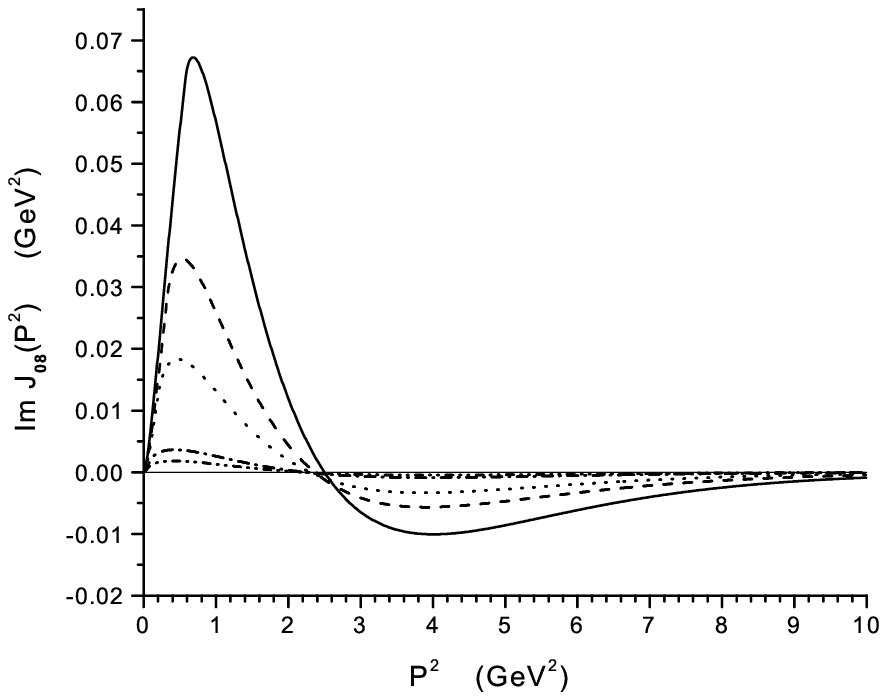}%
 \caption{Values of $\mbox{Im}\,J_{08}(P^2)$ are shown. [See the
 caption of Fig. 7.]}
 \end{figure}

 \begin{figure}
 \includegraphics[bb=0 0 280 230, angle=0, scale=1]{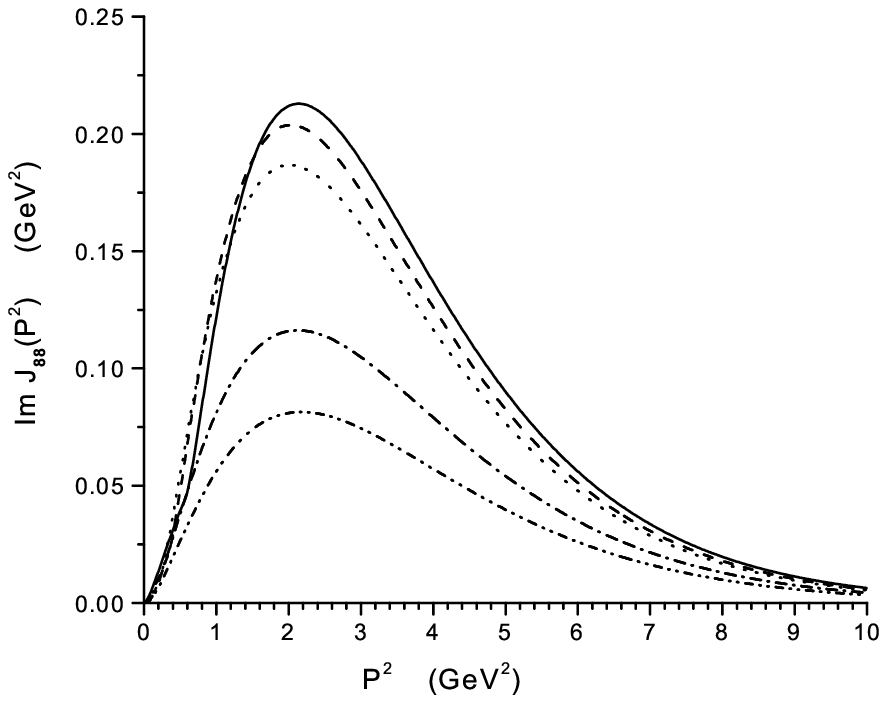}%
 \caption{Values of $\mbox{Im}\,J_{88}(P^2)$ are shown. [See the
 caption of Fig. 7.]}
 \end{figure}

In Figs. 10-12 we show $\mbox{Im}\,C_{00}(P^2)$,
$\mbox{Im}\,C_{08}(P^2)$ and $\mbox{Im}\,C_{88}(P^2)$. In the case
of $\mbox{Im}\,C_{00}(P^2)$ we see a sharp peak at 550 MeV. This
peak represents the lowest $f_0$ state at that temperature and has
as its ``parent" the $f_0(980)$ meson. [See Fig. 4] The decay of
the state at 550 MeV to two pions will be suppressed since the
pionic excitation see in Fig. 13 is essentially degenerate with
the $f_0$ excitation for $T>T_c$. (Note that in the flavor-SU(3)
version of the NJL model the $f_0(980)$ replaces the sigma of the
flavor-SU(2) NJL model as the chiral partner of the pion. Thus, it
is the state that evolves from the $f_0(980)$ with increasing
temperature that becomes degenerate with the pion upon restoration
of chiral symmetry.)

 \begin{figure}
 \includegraphics[bb=0 0 280 230, angle=0, scale=1]{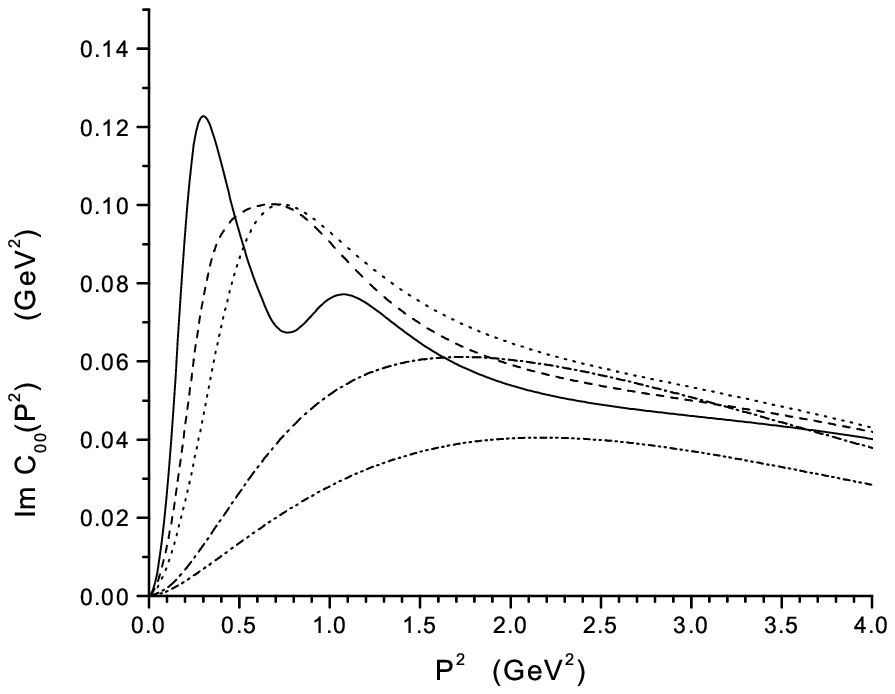}%
 \caption{Values of $\mbox{Im}\,C_{00}(P^2)$ are shown. [See the
 caption of Fig. 7.]}
 \end{figure}

 \begin{figure}
 \includegraphics[bb=0 0 280 230, angle=0, scale=1]{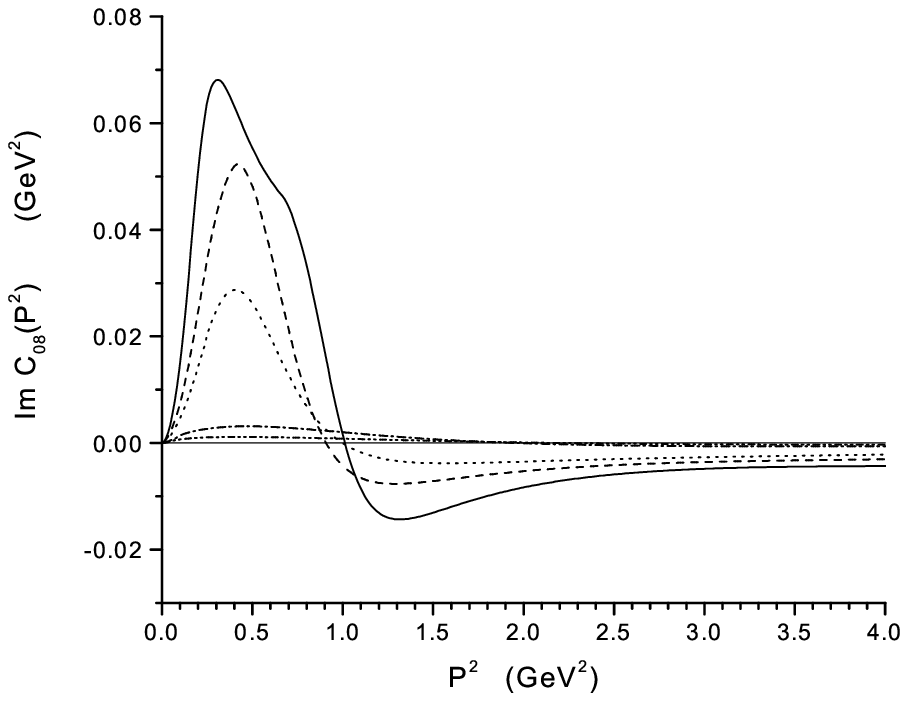}%
 \caption{Values of $\mbox{Im}\,C_{08}(P^2)$ are shown. [See the
 caption of Fig. 7.]}
 \end{figure}

 \begin{figure}
 \includegraphics[bb=0 0 280 230, angle=0, scale=1]{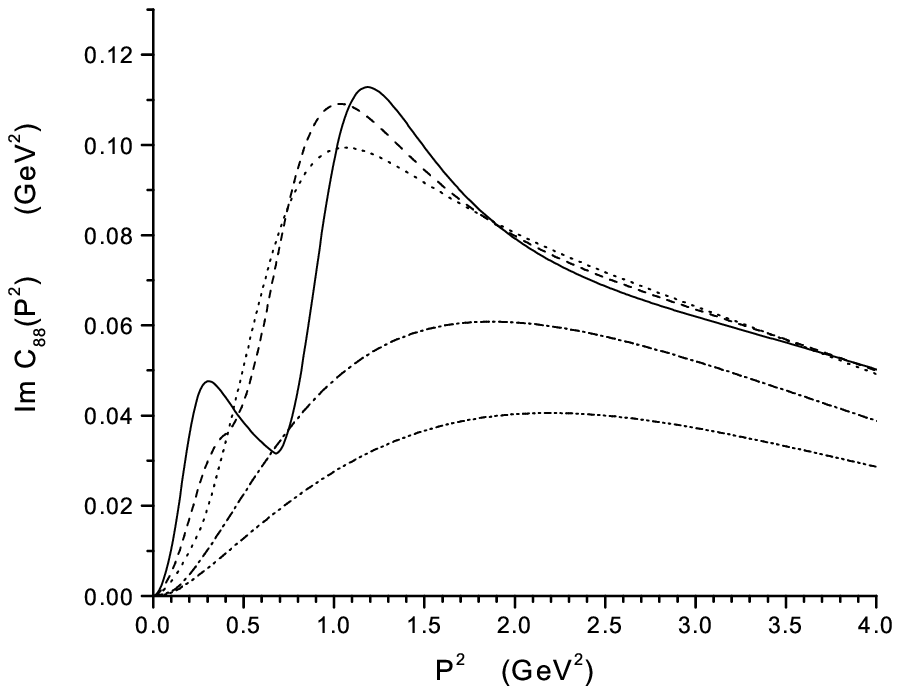}%
 \caption{Values of $\mbox{Im}\,C_{88}(P^2)$ are shown. [See the
 caption of Fig. 7.]}
 \end{figure}

 \begin{figure}
 \includegraphics[bb=0 0 280 230, angle=0, scale=1]{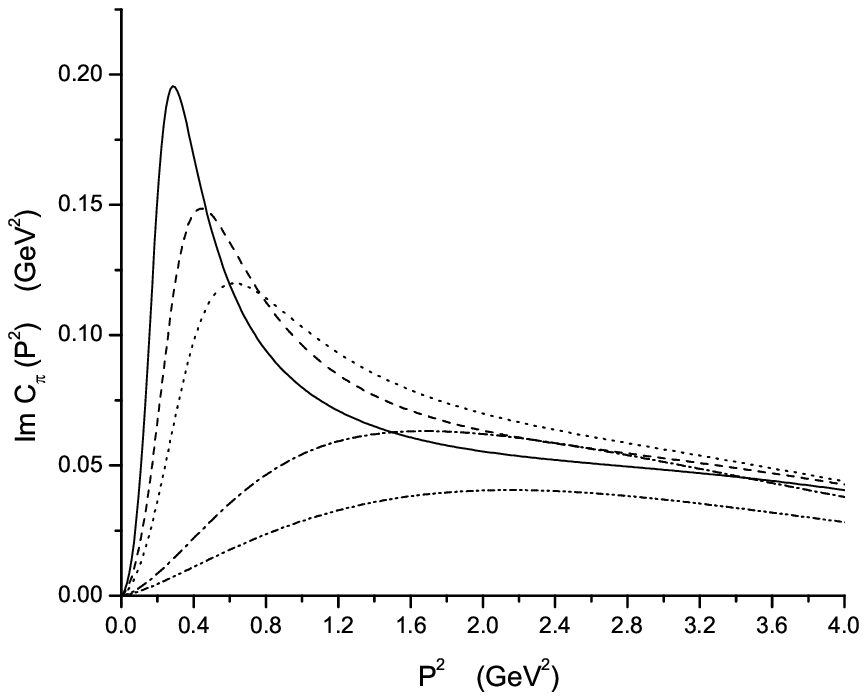}%
 \caption{Values of $\mbox{Im}\,C_{\pi}(P^2)$ are shown. [See the
 caption of Fig. 7.]}
 \end{figure}

The peak seen at about 1100 MeV in Fig. 12 is mainly an octet
state and is quite prominent in $\mbox{Im}\,C_{88}(P^2)$. When we
calculate the mixing angle for the state at 550 MeV, we find
$\theta_1=27.7^\circ$ corresponding to the state being about 78\%
flavor-singlet. A calculation of the mixing angle for the state at
1100 MeV yields $\theta_2=17.8^\circ$ which makes the state 95\%
flavor-octet. These mixing angles appear in the schematic
representation \be |f_0(247)\rangle=\cos
\theta_1\lambda_0+\sin\theta_1\lambda_8\,,\ee \be
|f_0(860)\rangle=-\sin
\theta_2\lambda_0+\cos\theta_2\lambda_8\,.\ee It may be seen in
Fig. 10 that the lowest $f_0$ state moves up in energy and becomes
wider when $T=1.6T_c$, while the state that is at 1100 MeV at
$T=1.2T_c$ moves down in energy at $T=1.6T_c$. (See Fig. 12.) We
also see that as the temperature is increased, the states broaden
further and the curves eventually become rather featureless.

In Fig. 13 we show $\mbox{Im}\,C_{\pi}(P^2)$ for the same set of
temperatures used to exhibit the properties of the
scalar-isoscalar excitations. The curve for $T=1.2T_c$ [solid
line] in Fig. 13 has a peak at 547 MeV. With increasing
temperature the excitation becomes much broader.

Results for $\mbox{Im}\,C_{\rho}(P^2)$ are shown in Fig. 14. There
is a peak at about 700 MeV that has as a ``parent" the
$\rho(770)$. Again, we see increased broadening with increasing
values of the temperature.

 \begin{figure}
 \includegraphics[bb=0 0 280 230, angle=0, scale=1]{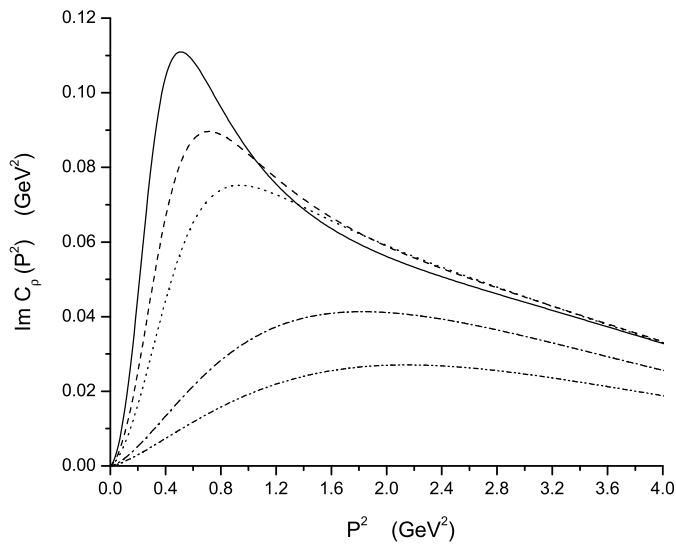}%
 \caption{Values of $\mbox{Im}\,C_{\rho}(P^2)$ are shown. [See the
 caption of Fig. 7.]}
 \end{figure}

\section{discussion}

The formalism we have developed allows for a description of the
mass values of various mesons and their radial excitations for the
range of temperatures $T<T_c$. Our model also describes the
confinement-deconfinement transition at $T=T_c$. When we
calculated hadronic current correlators in the real-time formalism
we are able to describe the widths of the excitations for $T>T_c$.
It is of interest to contrast our results with those obtained
using the imaginary-time formalism [4, 6, 22]. The results for the
behavior of the sigma and pion masses are similar in the NJL model
[4] and the field-theoretic model of Ref. [6]. In Ref. [6] the
value of $T_c\simeq150$ MeV is obtained when the current quark
mass is zero. For a finite value of the current quark mass, we may
make reference to Fig. 6 of Ref. [6]. There, the sigma mass has a
minimum value of approximately 255 MeV at $T=160$ MeV. At that
temperature the pion mass is about 220 MeV. If we consider
$T=1.2T_c=180$ MeV, we have $m_\pi=m_\sigma=440$ MeV in Ref. [6].
In our work, we have both $m_{f_0}$ and $m_\pi$ at about 550 MeV
when $T=1.2T_c$. Probably, one of the more significant differences
between the imaginary-time formalism used in Refs. [4] and [6] and
the real-time formalism used here is the absence of information
concerning the widths of the excitations in the imaginary-time
formalism. As may be seen in our Figs. 7-14, the widths of the
excitations may play a very important role in understanding the
properties of the quark-gluon plasma.

It is believed that measurements of particle ratios obtained in
central heavy-ion collisions contain useful information concerning
the properties of the quark-gluon plasma. In a recent study,
different assumptions concerning the dynamics of the plasma were
made and their influence upon predicted particle ratios were
obtained [24]. The authors consider calculations based upon both a
noninteracting gas model and a chiral SU(3) $\sigma-\omega$ model.
They find that ``the extracted chemical freeze-out parameters
differ considerably from those obtained in simple noninteracting
gas calculations..." when the $\sigma-\omega$ model is used to
calculate particle ratios. We believe our model may be useful in
performing calculations of the type reported in Ref. [24].


\vspace{1.5cm}
\noindent$\textbf{References}$\\[-2cm]


\end{document}